\documentclass[aps,prl,twocolumn,superscript,superscriptaddress,floatfix,citeautoscript,longbibliography]{revtex4-2}
\usepackage{relsize}
\usepackage{amsmath,amssymb} 
\usepackage{bm} 
\usepackage{graphicx} 
\usepackage{comment} 
\usepackage[normalem]{ulem} 

\usepackage{minted} 
\usepackage{textcomp} 

\usepackage{enumitem}
\setlist{noitemsep,leftmargin=*,topsep=0pt,parsep=0pt}

\usepackage{xcolor} 
\definecolor{lightgray}{gray}{0.6}
\definecolor{medgray}{gray}{0.4}
\definecolor{mRed}{RGB}{230, 0, 50}
\colorlet{newtextColor}{mRed}

\usepackage{hyperref}
\hypersetup{
colorlinks=true,
urlcolor= blue,
citecolor=blue,
linkcolor= blue,
}

\makeatletter
\renewcommand*{\fnum@figure}{\textbf{Fig.\,\thefigure}\,}
\renewcommand*{\@caption@fignum@sep}{\textbf{\,\textbar\,}}
\makeatother



\newif\ifptitle
\newif\ifpnumber
\newcounter{para}
\newcommand\ptitle[1]{\par\refstepcounter{para}
{\ifpnumber{\noindent\textcolor{lightgray}{\textbf{\thepara}}\indent}\fi}
{\ifptitle{\textbf{[{#1}]}}\fi}}

\makeatletter
\renewcommand\section{%
  \@startsection{section}{1}{\z@}%
  {1.5ex \@plus .5ex \@minus .2ex}
  {0.8ex \@plus .2ex}
  {\normalfont\large\bfseries\raggedright}%
}
\makeatother

\makeatletter
\renewcommand\subsection{%
  \@startsection{subsection}{2}{\z@}%
  {2ex \@plus .5ex \@minus .2ex}
  {0.01pt}
  {\normalfont\fontsize{11pt}{13pt}\bfseries\raggedright}%
}
\makeatother

\makeatletter
\renewcommand\subsubsection{%
  \@startsection{subsubsection}{3}{\z@}%
  {1.5ex \@plus .5ex \@minus .2ex}
  {0pt}
  {\normalfont\normalsize\bfseries\raggedright}%
}
\makeatother

\makeatletter
\def\@hangfrom@section#1#2#3{\@hangfrom{#1#2}#3}
\def\@hangfroms@section#1#2{#1#2}
\makeatother

\ptitlefalse  
\pnumberfalse 

\newcommand{\heng}{School of Engineering \& Applied Sciences, Harvard University, Cambridge, Massachusetts 02138, USA}
\newcommand{\hphys}{Department of Physics, Harvard University, Cambridge, Massachusetts 02138, USA}
\newcommand{\rice}{Department of Material Science and Nanoengineering, Rice University, Houston, TX, 77005, USA}

\newcommand{\LNb}{LiNbO$_3$}

\begin{document}

\title{A closed-loop platform for the design and nanoscale imaging \\ of GHz acoustic metamaterials}


\author{Federico Maccagno}
\author{Jasleen Kaur}
\author{Benjamin H. November}
\author{Layan Ansari}
\author{Daria-Teodora Harabor}
\author{Rares-Georgian Mihalcea}
\affiliation{\hphys}
\author{Harris Pirie}
\affiliation{\rice}
\author{Jennifer E. Hoffman}
\affiliation{\hphys}
\affiliation{\heng}

\date{\today}

\begin{abstract} 
Band structure engineering in surface acoustic wave (SAW) metamaterials could advance both classical telecommunications and quantum information processing. However, no imaging technique has demonstrated the necessary capability to resolve sub-$\mu$m traveling SAWs across wide GHz bandwidths. Existing methods capture only fragments of the dispersion at discrete frequencies, preventing systematic characterization and control of SAW-based metamaterials. Here, we develop electrostatic force microscopy (EFM) to enable real-space imaging of traveling SAWs in honeycomb metamaterials on \LNb. Our application leverages sub-200 nm spatial resolution, broad GHz bandwidth, and non-contact imaging to map complex band structures with continuous frequency resolution and expanded frequency range, while preserving sub-lattice detail. Using EFM, we map the full relevant frequency range around the Dirac point of a SAW graphene analog, including the acoustic Dirac cones, and the transition from ballistic to diffusive SAW transport regime. Furthermore, by breaking sublattice symmetry, we tune the opening of a band gap at the Dirac point, and image frequency-dependent wave localization on sublattice sites. Our EFM technique closes the loop between design and real-space validation, streamlining the engineering of arbitrary SAW landscapes for next-generation applications spanning telecommunications, microfluidics, and quantum acoustics.
\end{abstract}

\maketitle

\ptitle{Applications of \LNb}
The propagation of surface acoustic waves (SAWs) on piezoelectric materials provides a pathway for direct conversion between mechanical, electrical, and optical energy and information. SAWs have several advantages over other classical platforms for microwave applications. SAWs can easily be excited electrically up to tens of GHz, their micron-scale wavelengths allow for miniaturization, and their low losses in several piezoelectric substrates enable efficient transmission \cite{szaboITSU1973}. Nanofabrication of metamaterial lattices on the surface of LiNbO$_3$ can scatter traveling waves into standing waves, forming high quality resonators for optomechanical conversion or signal amplification \cite{mohammadiAPL2009, tadesseNC2014, zhuAOP2021, shaoPRA2019}. Alternatively, traveling SAWs can be coupled to superconducting qubits \cite{gustafssonNP2012, gustafssonS2014, satzingerN2018, chuS2017} or SiV spins \cite{MaityNatCom2020, XuPhysRevA2026} for quantum communication and information processing \cite{schuetzPRX2015}. However, no existing tool can directly image traveling GHz SAWs with high spatial resolution over a large continuous bandwidth, making it difficult to fine-tune the engineered lattices for diverse applications.

\ptitle{Intro to metamaterials/\LNb}
Metamaterials, whose wave behavior arises from their macroscopic geometry rather than the intrinsic properties of their microscopic building blocks, have been widely used to control wave dispersion in ways unattainable in pure materials. Acoustic metamaterials have enabled a wide range of engineered systems, including topological insulators \cite{bandresS2018, akbari-farahaniSaAAP2024, luNP2017, periS2020, niNM2019, weiNM2021, wangNC2022, serra-garciaN2018}, zero- or negative-index materials \cite{liLSA2021, duboisNC2017, tangNL2021}, and functional devices such as acoustic logic gates \cite{XiaAM2018, PiriePRL2022}. These applications exist primarily in the audible and ultrasonic regimes, corresponding to centimeter- to millimeter-scale wavelengths. The miniaturization of acoustic metamaterials to micron scales and microwave frequencies using SAWs presents challenges in both fabrication and measurement. Earlier efforts have implemented honeycomb lattices of deposited metallic micropillars to create SAW analogs of both the ``massless'' Dirac dispersion \cite{yuNM2016} and the topological features associated with sublattice-symmetry breaking \cite{wangNC2022, ZhangNatElec2022, zhangPRA2021}. However, previous characterization has been limited to transmission measurements, narrow MHz bandwidth, or discrete frequency imaging, leaving the full band structure and frequency-dependent scattering in the GHz regime unexplored. 
Existing techniques include microwave impedance microscopy (MIM) \cite{ZhangNatElec2022, shaoPRA2019, nii2023}, scanning interferometry \cite{Zanotto2022, Cha2018, wangNC2022, Yan2018}, and acoustic atomic force microscopy (AAFM) \cite{Pitanti2023}. However, each of these techniques faces challenges in recording fast, high-resolution, and large-bandwidth scans of traveling SAWs.
Optical readout techniques are bandwidth-limited both by the diffraction-limited focus spot, which sets the smallest imageable wavelength and thus the highest accessible SAW frequency, and by the photodetector speed, which bounds the frequency in a typical homodyne detection scheme.
Contact scanning techniques such as MIM and AAFM, while capable of higher spatial resolution, are constrained by the scan speed and the bandwidth of either the impedance-matching circuitry or cantilever-resonance demultiplexing. These limitations have so far prevented the reconstruction of key band structure features of GHz SAW metamaterials.

\begin{figure}[h!]
    \centering
    \includegraphics[width=1\columnwidth]
    {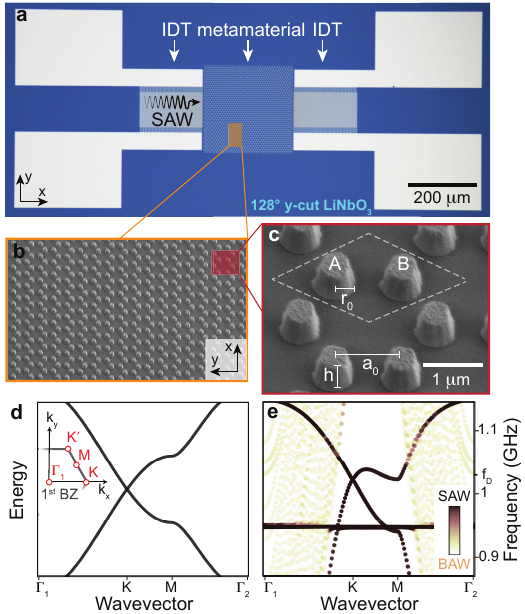}
    \caption{\textbf{Metamaterial design.} 
    \textbf{a}, Optical image of a single SAW metamaterial device. A lattice of gold pillars is deposited between two interdigital transducers (IDTs) used for SAW generation/detection in the 880-1150~MHz frequency range. 
    \textbf{b}, SEM image of the metamaterial (rotated by $90^{\circ}$). SAWs propagate along $\hat{x}$, corresponding to the $\Gamma-\text{K}$ direction of the honeycomb lattice.
    \textbf{c}, SEM image of the pillars, where the white rhombus indicates the unit cell, which consists of two gold nanopillars with $a_0 = 1.06\ \mu$m, $r_0 = 300$ nm, $h = 400$ nm.
    \textbf{d}, Tight-binding band structure of graphene showing the Dirac cone.
    \textbf{e}, Simulated acoustic band structure of the graphene metamaterial. The color scale represents the ratio of surface-to-bulk acoustic energy density, with darker markers corresponding to surface acoustic waves (SAWs) and lighter regions to bulk acoustic waves (BAWs). The flat bands around 950 MHz come from the radial expansion modes of the pillars which are decoupled from the substrate and do not appear in the experiment.}
\label{fig:metamaterial_design}
\end{figure}

\ptitle{Here we show...} Here we present electrostatic force microscopy (EFM) \cite{oliverAPL2001, hesjedalAPL2001, yahyaieITUFFC2012} as a tool for imaging the dispersion of GHz SAWs in \LNb\ metamaterials. EFM is an ideal platform because  
(i) it achieves sub-200 nm spatial resolution across a broad bandwidth (0.4-1.5 GHz);  
(ii) it is a non-contact technique, so it supports the fast scanning necessary for multi-frequency acquisition;  
(iii) it uses a heterodyne detection scheme, which eliminates the need for radiofrequency (RF) readout electronics or broadband impedance matching; and  
(iv) it can be implemented cost-effectively in most commercial atomic force microscope (AFM) setups. 
This combination permits complete mapping of GHz SAW metamaterial band structures for the first time. 
Using EFM, we scan a $\mu$m-scale \LNb\ metamaterial and visualize its acoustic Dirac cones in momentum space. We capture the transition from ballistic to diffusive transport, and image ``deaf'' bands that evade direct excitation \cite{yuNM2016, sanchez-perezPRL1998}. By fabricating a series of devices that break sublattice symmetry, we open a tunable band gap at the Dirac point and measure sublattice wave polarization, illustrating precise, on-chip control of SAW dispersion. Together, our results establish EFM as a powerful platform for characterizing acoustic metamaterials at GHz frequencies, opening new possibilities in microwave signal processing, acousto-electronic integration, and quantum information technologies.

\begin{figure*}[!t]
    \centering
    \includegraphics[width=\textwidth]{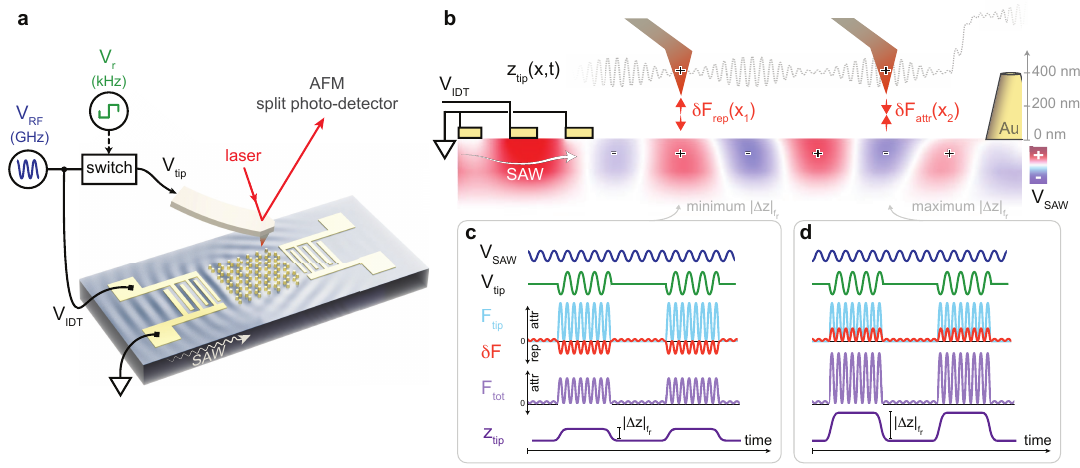}
\caption{
\textbf{Phase-resolved electrostatic force microscopy (EFM) of SAW metamaterials.}  
\textbf{a}, Schematic of the experimental setup, including the metamaterial device and the EFM used for detecting traveling SAWs. The EFM is a modified optical-detection AFM system in which the RF voltage used to drive the IDT is also applied to a conductive cantilever tip, with the tip signal chopped at the cantilever's mechanical resonance frequency $f_{\mathrm{r}}$.
\textbf{b}, Schematic of the tip–sample interaction at two different spatial locations, $x_1$ and $x_2$.
\textbf{c},\textbf{d}, Time evolution of the tip–sample interaction at $x = x_1$
where $V_{\mathrm{tip}}$ and $V_{\mathrm{SAW}}(x)$ are in-phase 
(\textbf{c}) and $x = x_2$
where $V_{\mathrm{tip}}$ and $V_{\mathrm{SAW}}(x)$ are $180^{\circ}$ out-of-phase 
(\textbf{d}). Both $V_{\mathrm{tip}}$ and $V_{\mathrm{IDT}}$ are driven by the same $V_{\mathrm{RF}}$; therefore each spatial location maintains its own constant tip-sample phase difference, conceptually akin to a path-difference interferometer. To understand the full tip-sample force during SAW imaging, we first consider the tip-sample force with no SAW (light blue), when $V_{\mathrm{IDT}}=0$, and the electrostatic interaction is driven entirely by $V_{\mathrm{tip}}$. In this case, the tip-sample force is always attractive, and proportional to $|V_{\mathrm{tip}}|$. When we turn on $V_{\mathrm{IDT}}$, the propagating SAW induces a weaker surface voltage $V_{\mathrm{SAW}}(x) \ll V_{\mathrm{tip}}$, producing a small additional tip-sample force contribution $\delta F(x)$ (dark orange). At locations such as $x=x_1$ (\textbf{c}),
where $V_{\mathrm{SAW}}(x)$ and $V_{\mathrm{tip}}$ are in-phase,
$\delta F(x)$ is maximally repulsive, reducing the total force $F_{\mathrm{tot}}(x)$ felt by the cantilever (purple). 
The cantilever responds primarily at its own mechanical resonance frequency (low-pass purple trace), resulting in minimum resonant oscillation amplitude $|\Delta z|_{f_\mathrm{r}}$. Conversely, at locations such as $x=x_2$ (\textbf{d}), 
where $V_{\mathrm{tip}}$ and $V_{\mathrm{SAW}}(x)$ are $180^{\circ}$ out-of-phase,
$\delta F(x)$ is maximally attractive, increasing the total force $F_{\mathrm{tot}}(x)$, and inducing maximal $|\Delta z|_{f_\mathrm{r}}$.}
\label{fig:EFM}
\end{figure*}

\ptitle{metamaterial design}
Our SAW metamaterial is composed of metallic pillars deposited in a honeycomb lattice on a piezoelectric $128^{\circ}$-Y cut bulk LiNbO$_3$ substrate. An overview optical image is shown in Fig.~\ref{fig:metamaterial_design}a, with higher-resolution scanning electron microscope (SEM) images provided in Fig.~\ref{fig:metamaterial_design}b,c. Each pillar acts as a local resonator coupled to the substrate, inducing periodic potential variations and thereby mimicking electronic dispersion in two-dimensional quantum materials \cite{Torrent2013, yuNM2016}. The pillar aspect ratio $h/r_0$ serves as the primary tuning knob for the resonance frequency of the pillar–substrate mode (see Extended Data Fig.~\ref{fig:mode_shape}), which plays the role of an onsite energy in a tight-binding analogy. The ratio of pillar radius to lattice spacing, $r_0/a_0$, controls the interpillar coupling, analogous to the hopping amplitude $t$. When the A and B sites (Fig.~\ref{fig:metamaterial_design}c) have equal radii, the structure mimics graphene (Fig.~\ref{fig:metamaterial_design}d), exhibiting Dirac degeneracies at the K and K$^\prime$ points of the Brillouin zone (BZ) (Fig.~\ref{fig:metamaterial_design}e). For our graphene metamaterial we use pillars with equal height ($h \approx 400$ nm), sidewall angle ($\theta \approx 18^{\circ}$), and base radius ($r_0 \approx 300$ nm), separated by a distance $a_0 \approx 1.06$ $\mu$m.  
Finally, we pattern two broadband chirped interdigital transducers (IDTs) to excite planar SAWs over the 850–1150~MHz range. The IDTs  generate waves in the \LNb\ $x$-crystallographic direction, as depicted in Fig.~\ref{fig:metamaterial_design}a. We fabricate the metamaterial and the IDTs in two separate lift-off steps using electron beam lithography and electron beam metal deposition (see Methods).

\begin{figure*}[!t]
    \includegraphics[width=\linewidth]{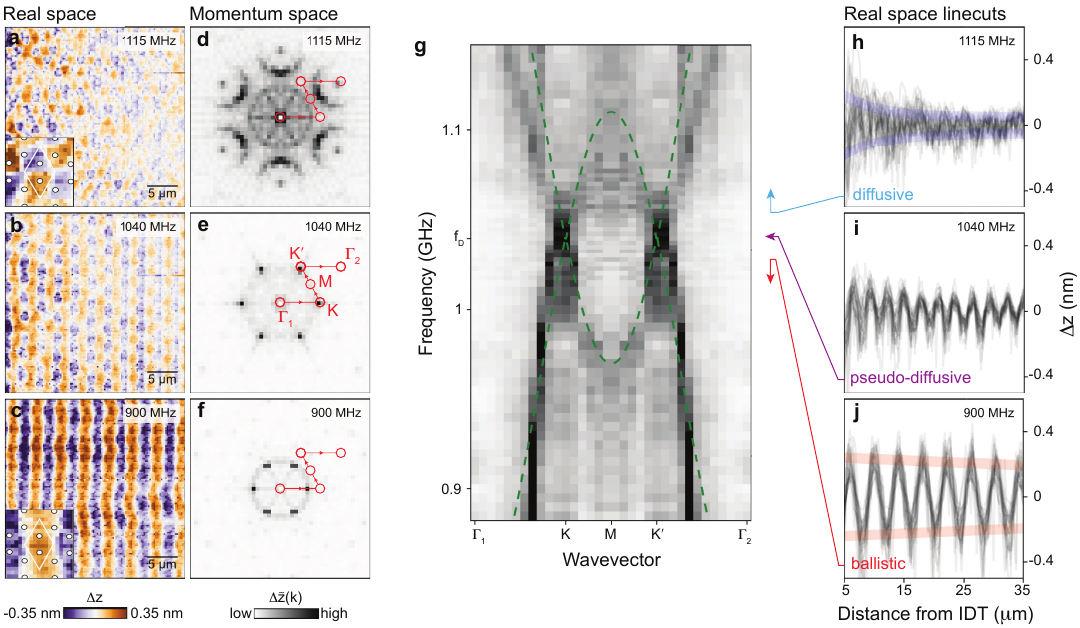}
\caption{\textbf{EFM observation of Dirac cones and transport regimes in a graphene-like SAW metamaterial.}
\textbf{a--c}, EFM amplitude shift $\Delta z$, measured at room temperature in air for frequencies above the Dirac frequency $f_{\mathrm{D}}$ (\textbf{a}), at $f_{\mathrm{D}}$ (\textbf{b}), and below $f_{\mathrm{D}}$ (\textbf{c}). 
A quadratic background was subtracted from each of the 128 scan lines for easier visualization of the spatial modulations.
Below $f_{\mathrm{D}}$ (\textbf{c}), the linear wavefronts are clearly visible, while at $f_{\mathrm{D}}$ (\textbf{b}) the waves form a triangular interference pattern. 
Above $f_{\mathrm{D}}$ (\textbf{a}), the amplitude drops because the mode symmetry suppresses direct excitation by the $\hat{x}$-propagating SAW, allowing only ``deaf'' modes driven by secondary scattering.
Insets in \textbf{a} and \textbf{c} show zoomed views of the unit cell, revealing antibonding (\textbf{a}) and bonding (\textbf{c}) modes. 
\textbf{d--f}, Six-fold symmetrized Fourier transforms of the real-space images in \textbf{a--c} (for raw FT data, see Extended Data Figs.~\ref{fig:graph_ext}-\ref{fig:gra_symm2}). 
Below $f_{\mathrm{D}}$ (\textbf{f}), a faint ring appears around $\Gamma_1$, with its intensity peaking along the $\Gamma$--K direction, corresponding to the unscattered plane waves. Both the ring and its peaks expand outward as the frequency approaches $f_{\mathrm{D}}$.
At $f_{\mathrm{D}}$ (\textbf{e}), sharp peaks localize at the K and K$^{\prime}$ corners of the BZ.
The modes above $f_{\mathrm{D}}$ are visible in (\textbf{d}) as circular bands closing around $\Gamma_2$, consistent with a graphene-like band structure.
\textbf{g}, Symmetrized band dispersion obtained from the $\Gamma_1$--K--M--K$^{\prime}$--$\Gamma_2$ normalized linecut versus frequency, showing Dirac cones at the K and K$^{\prime}$ points. 
The dashed green line is a tight-binding fit. 
\textbf{h--j}, Overlays of 32 distinct linecuts along the $\hat{x}$ direction from the real-space data in \textbf{a--c}. For $f < f_{\mathrm{D}}$ (\textbf{j}), transport is ballistic, with linear amplitude decay and coherent wavefronts. At $f_{\mathrm{D}}$ (\textbf{i}), we observe a crossover to a diffusive transport regime, where the amplitude decays exponentially with distance due to scattering; such diffusive behavior persists for all frequencies $f > f_{\mathrm{D}}$ (\textbf{h}).
}
    \label{fig:graphene}
\end{figure*}

\ptitle{EFM setup}  
During EFM imaging, we scan a platinum (Pt)-coated conductive cantilever $\sim 180$ nm above the sample surface (Fig.~\ref{fig:EFM}a), where the cantilever oscillations are driven solely by the electrostatic interaction with the underlying sample. We detect the cantilever position optically using the AFM's standard laser-photodiode system. We apply the same signal $V_{\mathrm{RF}}$ at GHz-range frequency $f_{\mathrm{RF}}$ to both the IDT (to excite the SAW) and the cantilever tip (to modulate the non-contact electrostatic interaction with the SAW). We chop the tip signal at the cantilever's kHz-range mechanical resonance frequency $f_{\mathrm{r}}$. 
The mechanical response of the tip effectively low-pass filters the GHz-range electrostatic force to a resonant kHz-range deflection.
In a simplified description, for a SAW of wavelength $\lambda$ propagating along the $x$-axis, the oscillation amplitude at $f_{\mathrm{r}}$ is given by \cite{yahyaieITUFFC2012}:
\begin{equation} 
|\Delta z|_{f_{\mathrm{r}}} = \frac{Q}{\pi k}
\biggl|
\underbrace{\frac{1}{2} \frac{\partial C_{\mathrm{ts}}}{\partial z} V_{\mathrm{tip}}\,^2
\vphantom{\left(\frac{2\pi x}{\lambda}\right)}}_{\displaystyle{F_{\mathrm{tip}}}}+
\underbrace{\frac{C_{\mathrm{ts}}\sigma_{0}}{\varepsilon} V_{\mathrm{tip}} \cos \left(\frac{2\pi x}{\lambda} \right)}_{\displaystyle{\delta F(x)}}
\biggr|_{f_{\mathrm{r}}}
\end{equation}
Here, $Q$ is the cantilever quality factor and $k$ the spring constant; $C_{\mathrm{ts}}$ is the tip–sample capacitance; $\sigma_0$ is the magnitude of the SAW-induced surface charge density; and $\varepsilon$ is the permittivity of air. The cantilever's resonant oscillation amplitude $|\Delta z|_{f_{\mathrm{r}}}$ is proportional to the magnitude of the electrostatic interaction force $|F_{z}|_{f_{\mathrm{r}}}$, i.e.\ the component of the total force oscillating at the cantilever's kHz mechanical resonance frequency $f_{\mathrm{r}}$, with negligible direct response to the GHz components. This resonant force contains two main contributions: a spatially-uniform attractive component, $F_{\mathrm{tip}}$, and a smaller perturbative component, $\delta F(x)$. The first term arises from the time-averaged electrostatic interaction between the tip and the sample, which can be modeled as a capacitor, where the tip is biased at $V_{\mathrm{RF}}$ and the sample is grounded. The resulting attractive force is obtained from the $z$-derivative of the electrostatic potential energy $U = \frac{1}{2} C_{\mathrm{ts}} V_{\mathrm{RF}}^2$.
The term $\delta F(x)$ depends on the SAW spatial phase ${2\pi x}/{\lambda}$, thus enabling direct imaging.
$\delta F(x)$ becomes maximally repulsive at positions where the tip voltage $V_{\mathrm{tip}}$ and the SAW-induced potential $V_{\mathrm{SAW}}(x)$ are in phase (e.g., $x = x_1$ in Fig.~\ref{fig:EFM}c). Conversely, $\delta F(x)$ is maximally attractive at positions where $V_{\mathrm{tip}}$ and $V_{\mathrm{SAW}}(x)$ are $180^{\circ}$ out of phase (e.g., $x = x_2$ in Fig.~\ref{fig:EFM}d).
Because $V_{\mathrm{tip}}$ and $V_{\mathrm{IDT}}$ are driven by the same $V_{\mathrm{RF}}$, the tip-sample phase relationship is fixed at each spatial location, conceptually akin to a path-difference interferometer. By measuring the cantilever oscillation amplitude $|\Delta z|_{f_{\mathrm{r}}}$ caused by the force $\delta F(x)$ as the tip moves over the sample, we can directly image the phase of the traveling SAW.

\begin{figure*}[!t]
    \includegraphics[width=\linewidth]{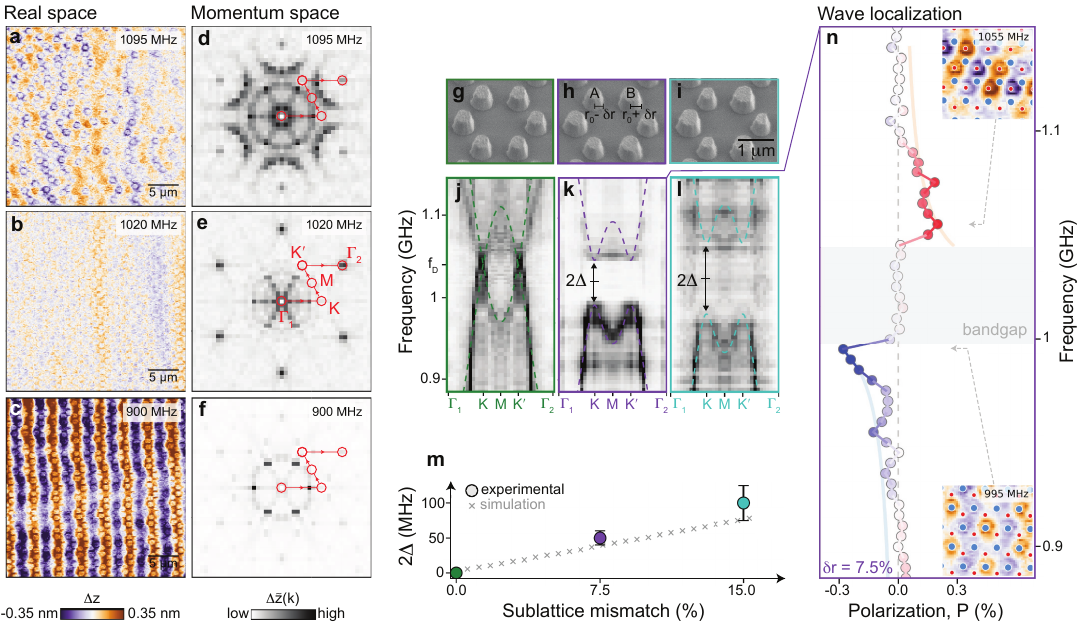}
\caption{\textbf{
EFM observation of tunable band gap and sublattice polarization in hBN-like SAW metamaterial.} 
\textbf{a--c,} EFM amplitude measured at room temperature in air for frequencies above (\textbf{a}), within (\textbf{b}), and below (\textbf{c}) the band gap in an hBN-like SAW metamaterial where the A and B sites have a radius mismatch of $\delta r = \pm7.5\%$. A quadratic background was subtracted from each of the 192 scan lines for easier visualization of the spatial modulations.
\textbf{d--f,} Six-fold symmetrized Fourier transforms of the images in \textbf{a--c} (for raw FT data, see Extended Data Figs.~\ref{fig:hbn_ext}-\ref{fig:hbn_symm2}). FTs show the same ballistic and diffusive transport regimes as the graphene-like metamaterial below and above the gap, but here the SAW is absent within the band gap (\textbf{b}) and induces no amplitude at the K point of the BZ (\textbf{e}). 
\textbf{g--i,} SEM images of honeycomb metamaterials with increasing sublattice asymmetry: $\delta r=0\%$ (\textbf{g}), 7.5\% (\textbf{h}), and 15\% (\textbf{i}). \textbf{j--l,} Reconstructed band structures measured on the metamaterials in \textbf{g--i}.
The Dirac frequency $f_{\mathrm{D}}$ of our hBN-like metamaterial is located at 1020~MHz, slightly lower than that of the graphene analog at 1040~MHz, due to fabrication tolerances. 
In \textbf{k} and \textbf{l}, band gaps of $\Delta \approx 50$~MHz and 100~MHz open around 1020~MHz due to sublattice symmetry breaking, while two dispersive bands remain visible above and below the gaps. Note that we did not normalize panels \textbf{k} and \textbf{l} at each frequency, so the gap remains visible in the raw data. 
\textbf{m,} Experimental and simulated data showing that the SAW band gap $\Delta$ scales with the fabricated sublattice mismatch $\delta r$. 
\textbf{n,} Sublattice polarization ratio $P = (|\Delta z_{\mathrm{A}}| - |\Delta z_{\mathrm{B}}|)\,/\,(|\Delta z_{\mathrm{A}}| + |\Delta z_{\mathrm{B}}|)$ between A (red dots in inset) and B (blue dots) sites reveals how SAWs localize near the band gap. As the frequency approaches the band gap from below, the SAW localizes on the larger B pillars with maximum effect at 995~MHz (lower inset). Approaching the band gap from above, localization shifts to the smaller A pillars, peaking at 1055~MHz (upper inset). Within the band gap, the overall signal drops and $P$ approaches zero.}
\label{fig:hbn}
\end{figure*}

\ptitle{Graphene metamaterial band structure}  
EFM images of our graphene-like metamaterial reveal three distinct transport regimes (Fig.~\ref{fig:graphene}). Below the Dirac frequency $f_{\mathrm{D}}$ (1040~MHz), the SAW propagates in coherent wavefronts along the excitation direction (Fig.~\ref{fig:graphene}c), appearing as sharp peaks along the $\Gamma_1$--K direction in the first BZ, with only minor scattering into states forming a faint circular contour around $\Gamma_1$ (Fig.~\ref{fig:graphene}f). Both the peaks and the circular contour move outward as the frequency approaches $f_{\mathrm{D}}$. At $f_{\mathrm{D}}$, triangular interference patterns emerge along the high-symmetry directions of the metamaterial (Fig.~\ref{fig:graphene}b), corresponding to sharp peaks at the K and K$^{\prime}$ corners of the first BZ (Fig.~\ref{fig:graphene}e). Above $f_{\mathrm{D}}$, the SAW amplitude weakens, indicating strong scattering (Fig.~\ref{fig:graphene}a). In momentum space, these scattered states appear around the $\Gamma_2$ Bragg peaks as ring-like features, which progressively shrink as the frequency is further increased (Fig.~\ref{fig:graphene}d).
We reconstruct the continuous band structure along $\Gamma_1$--K--M--K$^{\prime}$--$\Gamma_2$ from momentum maps at 66 frequencies, revealing linearly dispersing Dirac cones at K and K$^{\prime}$ (Fig.~\ref{fig:graphene}g). 
We measure the group velocity of 2730 m/s at the Dirac cone, consistent with the simulated value of 2830 m/s, but $\sim 30\%$
slower than free SAWs on \LNb\ \cite{Morgan2007SAW}. Such velocity engineering highlights the versatility of structured lattices for compressing the wavelength and controlling the dispersion of SAWs at GHz frequencies.

\ptitle{Ballistic-diffusive transition} 
An additional advantage of our EFM technique is its ability to directly image how the SAW decays with distance, enabling a clear distinction between ballistic and diffusive transport regimes. For frequencies below $f_{\mathrm{D}}$, the EFM amplitude decreases linearly while maintaining phase coherence (Fig.~\ref{fig:graphene}j), consistent with ballistic propagation and minimal scattering. Near $f_{\mathrm{D}}$, the decay is more complex, entering a pseudo-diffusive regime where the scattered energy flux from each pillar is omnidirectional, and the group velocity direction becomes ill-defined \cite{yuNM2016}. Phase coherence of the wave persists, but the amplitude decays more quickly than expected for ballistic propagation (Fig.~\ref{fig:graphene}i). Above $f_{\mathrm{D}}$, the amplitude decays exponentially and phase coherence is lost, indicating strong scattering in the diffusive regime (Fig.~\ref{fig:graphene}h).

\ptitle{``Deaf'' modes}  
Deaf modes dominate the SAW response above $f_{\mathrm{D}}$, producing a sharp reduction in conventional transport measurements \cite{yuNM2016}. Exciting a SAW mode requires matching both the frequency and spatial symmetry of the mode. Below $f_{\mathrm{D}}$, the two pillars within the rhombic unit cell oscillate in phase, forming symmetric modes that couple efficiently to the plane wave launched by the IDT (Fig.~\ref{fig:graphene}c, inset). However, above $f_{\mathrm{D}}$, the pillars oscillate out of phase, forming antisymmetric modes with a central node (Fig.~\ref{fig:graphene}a, inset). Because the IDT-driven plane wave cannot project onto these anti-symmetric eigenstates, they appear as an effective band gap. These so-called ``deaf'' bands \cite{sanchez-perezPRL1998} have prevented access to the dispersion of modes above $f_{\mathrm{D}}$ in previous transport experiments. We overcome this limitation because secondary scattering processes within the metamaterial can excite the anti-symmetric mode, generating a local response. The near-field sensitivity of EFM allows us to detect these waves locally (Fig.~\ref{fig:graphene}a), enabling reconstruction of the upper-band dispersion (Fig.~\ref{fig:graphene}g).

\ptitle{hBN band gap}  
Introducing a difference $\delta r$ in the pillar radii between sublattice sites A and B reduces the lattice symmetry from C$_6$ to C$_3$, opening a gap at the Dirac point analogous to the band structure of hexagonal boron nitride (hBN).
Such symmetry-tuned band gaps could host topological states that enable robust, defect-tolerant SAW waveguides \cite{ZhangNatElec2022,wangNC2022,Yan2018,nii2023}, facilitating new acoustic and acousto-electronic technologies \cite{PiriePRL2022}. 
Our EFM measurements on the hBN-like lattice directly reveal the formation of the predicted band gap (Fig.~\ref{fig:hbn}). With $\delta r = 7.5\%$, the behavior away from $f_{\mathrm{D}}$ remains similar to that of the graphene-like lattice, showing the transition from ballistic transport below $f_{\mathrm{D}}$ to diffusive transport above $f_{\mathrm{D}}$ (Fig.~\ref{fig:hbn}a,c,d,f). Near $f_{\mathrm{D}}$, however, a band gap opens, with no visible propagating waves (Fig.~\ref{fig:hbn}b). The corresponding Fourier transform (Fig.~\ref{fig:hbn}e) shows only pronounced Bragg peaks, indicating that most of the EFM signal arises from imaging the underlying pillars rather than any SAW. The reconstructed band structure along the $\Gamma_1$--K--M--K$^{\prime}$--$\Gamma_2$ path (Fig.~\ref{fig:hbn}k) reveals a $\sim50$ MHz gap, within which no propagating SAW modes exist.
Consistent with this picture, the area-averaged EFM signal drops below the noise floor ($–20$ dB) as the frequency is increased from the lower band (Fig.~\ref{fig:hbn}c) into the gap (Fig.~\ref{fig:hbn}b), before recovering to a detectable level of approximately $–10$ dB upon entering the upper band (Fig.~\ref{fig:hbn}a), where propagating ``deaf'' modes can still be excited via secondary scattering.
The difference in pillar radius serves as a tuning knob for the gap (Fig.~\ref{fig:hbn}m). When the sublattice symmetry breaking is increased from $\delta r = 7.5\%$ to $15\%$, the gap approximately doubles from $\sim 50$ MHz to $\sim 100$ MHz. This near-linear trend agrees well with our COMSOL simulations (Fig.~\ref{fig:hbn}m).

\ptitle{hBN sublattice polarization}  
Near the band gap edges, the SAW exhibits sublattice polarization.
We quantify this effect as a function of frequency in Fig.~\ref{fig:hbn}n by averaging the local EFM signal $|\Delta z|$ within a radius of $\sim 200$ nm around each pillar site in the scan area, and comparing its amplitude on A and B sites using an empirical polarization metric 
$P = (|\Delta z_{\mathrm{A}}| - |\Delta z_{\mathrm{B}}|)\,/\,(|\Delta z_{\mathrm{A}}| + |\Delta z_{\mathrm{B}}|)$. As the frequency approaches the band gap from below, we observe wave extrema increasingly localized on the larger B pillars, reaching a peak negative polarization $P = -0.28$ at 995~MHz, as shown in the lower inset of Fig.~\ref{fig:hbn}n.
At this frequency, corresponding to the upper edge of the lower band, the wave forms a triangular real-space pattern, corresponding to the same six $k$-space peaks at K and K$^{\prime}$ as seen in graphene at $f_{\mathrm{D}}$. Conversely, approaching the gap from above, we observe wave extrema increasingly localized on the A sites, with maximum polarization $P=+0.20$ at 1055 MHz, as shown in the upper inset of Fig.~\ref{fig:hbn}n. Within the band gap, We observe the overall signal magnitude decrease, and $P$ approaches zero. This behavior is qualitatively captured by a massive Dirac Hamiltonian for acoustic Bloch modes (see `Polarization from Dirac equation' in Methods). For this Hamiltonian, the sublattice polarization outside the gap is expected to follow
$P(f) = \alpha \Delta/(f - f_{\mathrm{D}})$, where in an ideal model $\alpha = 1$.
Experimentally, we find that a value of $\alpha = 0.28$ best fits the data for $\Delta = 25$ MHz and $f_{\mathrm{D}} = 1020$ MHz, as shown in Fig.~\ref{fig:hbn}n. The reduced polarization relative to the ideal case arises from several experimental factors: finite uncertainties in site localization and area averaging, broadening of the EFM signal due to pillar-induced field distortion and a sensing height of 180 nm, and incomplete coupling efficiency between the metamaterial and the propagating SAW.

\ptitle{Conclusion}  
The development of SAW metamaterials has been limited by the absence of an microscopy technique capable of imaging phase-resolved traveling waves across their GHz band structure. In this work, we establish electrostatic force microscopy (EFM) as a powerful probe of traveling surface acoustic waves in structured piezoelectric substrates, enabling wide-band operation at GHz frequencies with $\lesssim 200$ nm spatial resolution. Using a honeycomb array of Au pillars on \LNb, we directly image the real-space signatures of acoustic Dirac physics, including the crossover from ballistic to diffusive propagation and the presence of ``deaf'' modes that remain invisible to conventional far-field transport methods. Our capability opens a route to resolving the full acoustic band dispersion of SAW metamaterials while retaining access to sublattice spatial resolution. By extending the platform to an hBN analog with broken sublattice symmetry, we observed the opening of a tunable band gap, as well as strong sublattice-selective localization of the wave amplitude at the band edges. Together, these results establish a versatile framework for probing acoustic metamaterials with high spatial resolution, sensitivity and large bandwidth. The ability to visualize and quantify both propagating and evanescent modes closes the design loop for next-generation technologies such as GHz topological waveguides, sub-wavelength acoustic devices, and hybrid phononic–electronic systems where accurate control of energy flow at the nanoscale is essential.

\clearpage

\vspace{5mm}
\section*{Methods}

\subsubsection*{Metamaterial design and fabrication.\ }
SAW metamaterials were fabricated on a 600 $\mu$m thick, $128^{\circ}$-Y cut bulk \LNb\ substrate (Precision Micro-Optics). Gold pillars (with 5 nm chromium adhesion layer) were deposited by e-beam evaporation in a lift-off process, using a PMMA/MMA bilayer resist mask patterned using electron beam lithography (Elionix HS50). The pillars have height $h \sim 400$ nm and are arranged with separation $a = 1.060$ $\mu$m to set the Dirac point near 1.025 GHz. Interdigital transducers (IDTs) with a 300 MHz bandwidth were fabricated on both sides of the pillar lattice, using 5 nm Cr and 75 nm Au in a second lift-off step. Each IDT comprised 49 signal-ground finger pairs with periods decreasing from 5.48 $\mu$m to 3.81 $\mu$m. The IDTs were not impedance-matched, so the piezoelectric transduction efficiency is expected to be relatively low.

\subsubsection*{Finite Element Simulation.\ }
The metamaterial dispersion was simulated using a rhombohedral unit cell with Floquet periodic boundary conditions in COMSOL Multiphysics. A bulk \LNb\ thickness of 40 $\mu$m was used in the model. The elasticity, piezoelectric coupling, and permittivity matrices describing single-crystal \LNb \cite{AndrushchakJAP2009} were rotated using Euler angles $(0,38^\circ,0)$ to represent the intended 128$^\circ$ Y-cut \LNb\ substrate \cite{Morgan2007SAW}. Surface acoustic wave (SAW) modes were identified by comparing the elastic energy within the top 2.6~$\mu$m region nearest the surface $(-2.6~\mu\mathrm{m} < z < 0)$, to the energy integrated over the entire unit-cell volume $(-40~\mu\mathrm{m} < z < 0)$.

\subsubsection*{Electrostatic force microscopy.\ }
EFM measurements were performed on an Asylum Research Cypher~S AFM microscope \cite{CypherAFM} using conductive cantilevers (OPUS~240AC-PP) with resonance frequency $f_{\mathrm{r}} \approx 70$ kHz and stiffness $k \approx 2$ N/m \cite{MikroMasch_OPUS}. An Agilent E5062A vector network analyzer (VNA) provided the RF signal (power amplitude optimized for signal-to-noise: $-0.4$ dBm for graphene, $-1.5$ dBm for hBN 7.5\%, and $-0.5$ dBm for hBN 15\%). The VNA output was amplified by 22 dB using a Mini-Circuits ZX60-14LN-S+ amplifier. The amplified signal was applied to both the IDT (to generate SAWs) and the conductive tip holder through an SMA~T connector, where it was modulated at the cantilever resonance frequency $f_{\mathrm{r}}$ using a Mini-Circuits ZYSWA-2-50DR+ switch.
The cantilever motion was detected optically, and the resonant oscillation amplitude was demodulated using the Cypher~S controller's internal lock-in amplifier. 
The scanning procedure consisted of two phases: each scan line was recorded first in AC tapping mode to obtain the surface topography, and second in non-contact lift-mode. In the lift-mode scan, the cantilever retraces the measured topography with a constant offset height. During the lift-mode measurement, no mechanical excitation was applied to the cantilever; the oscillation was driven solely by electrostatic tip-sample forces. Lift heights ranging from 50 nm to 400 nm produced good results. We chose lift height 180~nm to optimize the signal-to-noise ratio while providing sufficient clearance to avoid tip–pillar collisions and snap-in-contact instabilities. This 180~nm distance refers only to the programmed lift \textit{offset} and does not include the average cantilever–surface separation during the first tapping-mode pass, which is estimated to be $\sim10$ nm. 
Typical scan parameters included 128--192 lines per frame, recorded at 0.5--2~s per line. A typical scan area of 30 $\mu$m $\times$ 30 $\mu$m with a lateral pixel spacing of 230~nm was acquired in approximately 2 minutes. The spatial resolution of EFM was estimated to be on the order of the tip–sample distance \cite{yahyaieITUFFC2012}. 
The images of the graphene metamaterial were acquired at a lateral distance of $\sim5$ $\mu$m from the source IDT, whereas the images of the hBN metamaterials were each acquired at a lateral distance of $\sim25$ $\mu$m from their IDTs. The graphene dataset consists of frequency scans acquired at 5~MHz intervals from 885 to 1150\,MHz, except in the range 1015--1080 MHz, where scans were acquired every 2.5~MHz. The hBN datasets were acquired at uniform 5~MHz intervals. All measurements were performed at ambient conditions, where the cantilever quality factor is strongly reduced ($\sim10^2$ in air vs.\ $\sim10^4$ in vacuum) demonstrating that this technique can be  implemented cheaply and effectively in commercial AFM systems while retaining good signal-to-noise ratio.

\subsubsection*{Polarization from Dirac equation.\ }
The dispersion for the acoustic Bloch modes near the K and K$^\prime$ points can be approximated as
$f(\mathbf{q}) = v\,\mathbf{q}\cdot\boldsymbol{\sigma} + \Delta \,\sigma_z$, where $v$ is the Dirac velocity and $2\Delta$ is the bandgap.
The eigenfrequencies can be computed as
$f_\pm(\mathbf{q}) = f_{\mathrm{D}} \pm \sqrt{v^2 |\mathbf{q}|^2 + \Delta^2}$. The corresponding eigenvectors can be written as
$\psi_\pm = \bigl(\pm\cos\frac{\theta}{2},\,\sin\frac{\theta}{2}\bigr)$ for the top ($+$) and bottom ($-$) bands, where
$\cos\theta = \Delta/\sqrt{v^2 |\mathbf{q}|^2 + \Delta^2}$. Here we are using a two-component spinor
$\psi = (u_{\mathrm{A}}, u_{\mathrm{B}})$,
where $|u_{\mathrm{A}}|^2$ and $|u_{\mathrm{B}}|^2$ denote the acoustic displacement amplitudes
on sites A and B. Define the acoustic sublattice polarization
\begin{equation}
P \equiv \frac{|u_{\mathrm{A}}|^2 - |u_{\mathrm{B}}|^2}{|u_{\mathrm{A}}|^2 + |u_{\mathrm{B}}|^2}.
\label{eq:polarization_u}
\end{equation}
Using the eigenvectors $\psi_\pm$,
\begin{equation}
P_\pm(\mathbf{q})
= \pm \frac{\Delta}{\sqrt{v^2 |\mathbf{q}|^2 + \Delta^2}}\,,
\label{eq:polarization_q}
\end{equation}
which can be written as
$P(f) = \Delta/(f - f_{\mathrm{D}})$
for frequencies outside the gap, $|f - f_{\mathrm{D}}| \ge \Delta$. As $f \to f_{\mathrm{D}} - \Delta$ (lower band edge), the polarization approaches $P \to -1$, indicating that the mode becomes predominantly B-like. Conversely, as $f \to f_{\mathrm{D}} + \Delta$ (upper band edge), the polarization approaches $P \to +1$, corresponding to a mode that is predominantly A-like. Far from the gap, $P \to 0$. 

\section*{Acknowledgments}
Work was supported by the Center for Integrated Quantum Materials under National Science Foundation Grant No.\ DMR-1231319. Fabrication and imaging were performed in the Harvard University Center for Nanoscale Systems (CNS), a member of the National Nanotechnology Coordinated Infrastructure Network (NNCI), which is supported by the National Science Foundation under NSF award no.\ ECCS-2025158.

\section*{Author Contributions}
F.M., B.H.N., H.P., and J.E.H. conceived the experiment. F.M. simulated and designed the devices. F.M., L.A., J.K., D.T.H. and R.G.M. developed the fabrication process. F.M. fabricated the devices and built and operated the EFM setup. F.M. and J.K. acquired the data for the graphene and 7.5\% hBN devices, while J.K. and D.T.H. acquired the data for the 15\% hBN device. F.M., H.P. and B.H.N. carried out the data analysis. F.M., B.H.N., H.P., and J.E.H. wrote the manuscript.

\clearpage\onecolumngrid
\section*{Extended Data}

\setcounter{figure}{0}
\makeatletter
\renewcommand*{\fnum@figure}{\textbf{Extended Data Fig.\,\thefigure}\,}
\renewcommand*{\@caption@fignum@sep}{\textbf{\,\textbar\,}}
\makeatother


\begin{figure*}[h]
    \includegraphics[width=\linewidth]{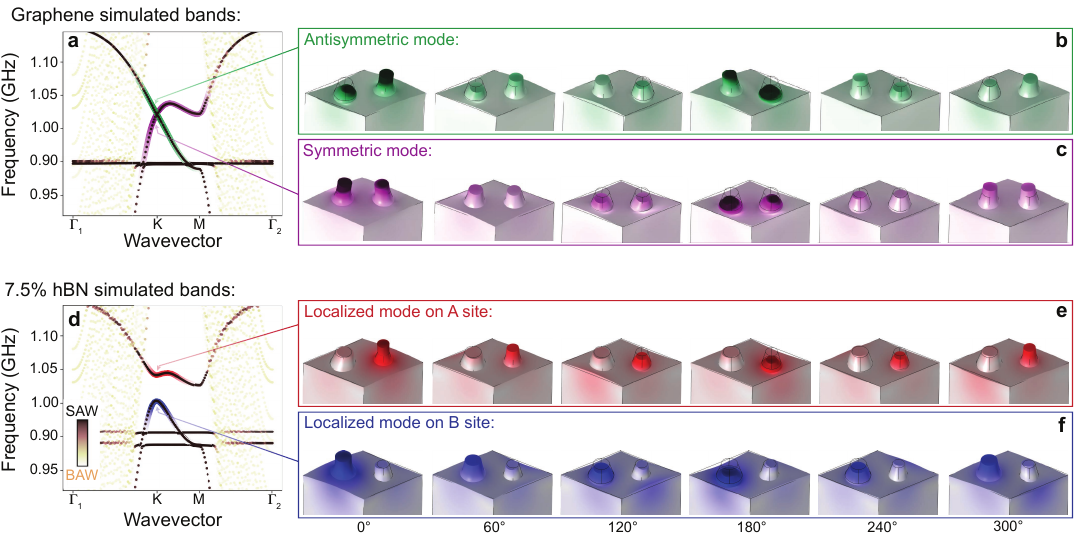}
\caption{\textbf{COMSOL Multiphysics acoustic mode simulations.}
\textbf{a,} Simulated acoustic band structure of the graphene metamaterial. The color scale represents the ratio of surface-to-bulk acoustic energy density, with darker markers corresponding to surface acoustic waves (SAWs) and lighter markers corresponding to bulk acoustic waves (BAWs). The purple dispersive band corresponds to the mode in which the pillars move out of plane \emph{in phase}, which we refer to as the symmetric or ``bonding'' mode. This motion induces equal pressure and therefore equal piezoelectric response in the substrate beneath the pillars, producing no node between pillars in the EFM signal (see Fig.~\ref{fig:graphene}\textbf{c} inset). The green band corresponds to the mode in which the pillars move out of plane with a $180^{\circ}$ phase difference. This creates high pressure under one pillar and low pressure under the other, generating opposite piezoelectric responses and a distinct node in the EFM signal (see Fig.~\ref{fig:graphene}\textbf{a} inset). We refer to this as the antisymmetric or ``antibonding'' mode.
\textbf{b--c,} Evolution of the mode shape across a phase cycle for the antisymmetric (\textbf{b}) and symmetric (\textbf{c}) modes. Simulations are shown at $k \approx \mathrm{K}$. The color intensity represents the displacement magnitude relative to the rest position.
\textbf{d,} Simulated acoustic band structure of the hBN ($\delta r = 7.5\%$) metamaterial. The modes at the lower edge of the upper band (red) and the upper edge of the lower band (blue) show clear sublattice localization.
\textbf{e--f,} Evolution of the mode shape across a phase cycle for the upper-band-edge mode (\textbf{e}) and lower-band-edge mode (\textbf{f}), shown at $k = \mathrm{K}$. In the upper-band-edge mode the participation localizes on the smaller pillar A, whereas in the lower-band-edge mode, the participation is dominated by the larger pillar B. This behavior is consistent with our EFM observations (see Fig.~\ref{fig:hbn}\textbf{k}).
}
\label{fig:mode_shape}
\end{figure*}

\newpage

\begin{figure*}[!t]
    \includegraphics[width=\linewidth]{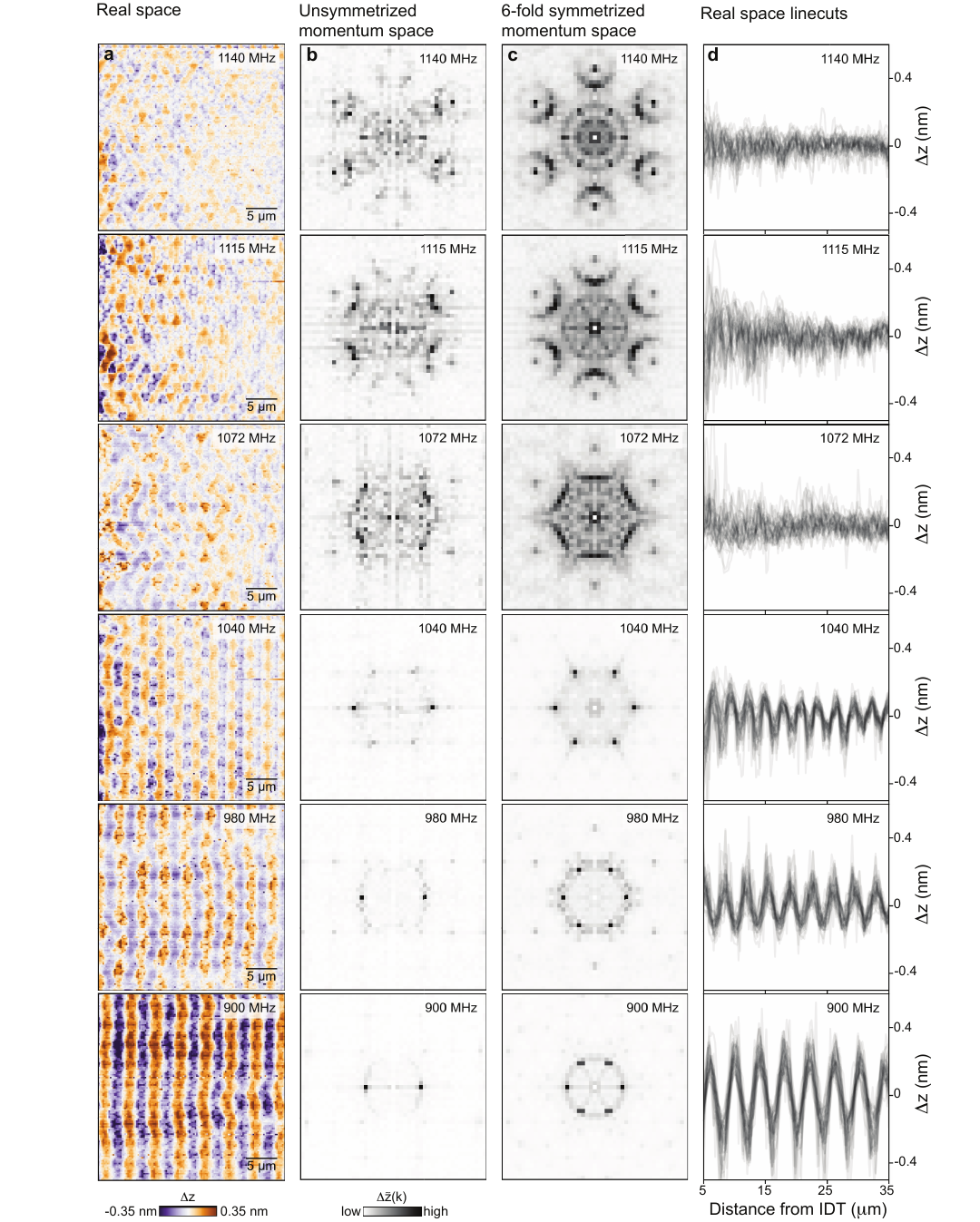}
\caption{
\textbf{EFM characterization of the graphene metamaterial.} Same data series as in Fig.~\ref{fig:graphene}, but showing additional frequencies and analysis steps.
\textbf{a,} EFM amplitude shift $\Delta z$ measured at room temperature and ambient pressure for the graphene metamaterial. 
A quadratic background was subtracted from each of the 128 scan lines. 
\textbf{b,} Raw Fourier transform of the data in (\textbf{a}). 
\textbf{c,} Six-fold symmetrized data derived from (\textbf{b}). 
\textbf{d,} Overlays of 32 distinct linecuts along the $\hat{x}$ direction from the real-space data in (\textbf{a}), equally spaced in $\hat{y}$.}
\label{fig:graph_ext}
\end{figure*}

\newpage

\begin{figure*}[h]
    \includegraphics[width=\linewidth]{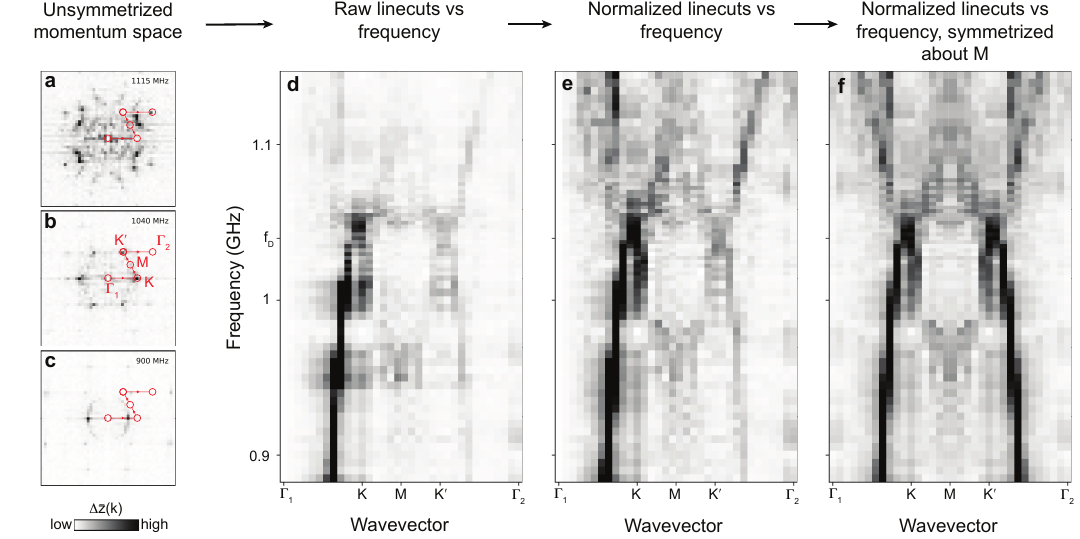}
\caption{\textbf{Graphene acoustic bands from non-symmetrized Fourier-space data.}
\textbf{a--c,} Raw Fourier transforms for the graphene metamaterial measured above the Dirac frequency $f_{\mathrm{D}}$ (\textbf{a}), at $f_{\mathrm{D}}$ (\textbf{b}), and below $f_{\mathrm{D}}$ (\textbf{c}). 
\textbf{d,} Band dispersion obtained from the $\Gamma_{1}$--K--M--K$^{\prime}$--$\Gamma_{2}$ linecut in momentum space (shown as a dashed red line in (\textbf{b}))  using the data in (\textbf{a}--\textbf{c}). 
\textbf{e,} Band dispersion obtained by normalizing each frequency linecut in (\textbf{d}) about its mean. 
\textbf{f,} Band dispersion obtained after symmetrizing the data from (\textbf{e}) about the M point.}
\label{fig:gra_symm1}
\end{figure*}

\begin{figure*}[!t]
    \includegraphics[width=\linewidth]{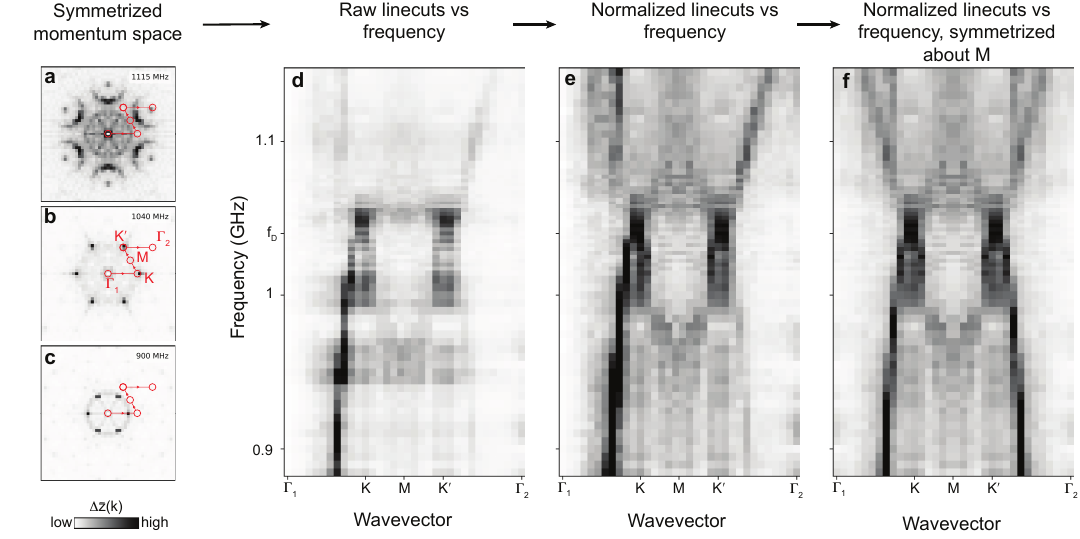}
\caption{\textbf{Graphene acoustic bands from symmetrized Fourier-space data.}
\textbf{a--c,} Six-fold symmetrized Fourier transforms for the graphene metamaterial measured above the Dirac frequency $f_{\mathrm{D}}$ (\textbf{a}), at $f_{\mathrm{D}}$ (\textbf{b}), and below $f_{\mathrm{D}}$ (\textbf{c}). 
\textbf{d,} Band dispersion obtained from the $\Gamma_{1}$--K--M--K$^{\prime}$--$\Gamma_{2}$ linecut in momentum space (shown as a dashed red line in (\textbf{b}))  using the data in (\textbf{a}--\textbf{c}). 
\textbf{e,} Band dispersion obtained by normalizing each frequency linecut in (\textbf{d}) about its mean. 
\textbf{f,} Band dispersion obtained after symmetrizing the data from (\textbf{e}) about the M point.}
\label{fig:gra_symm2}
\end{figure*}

\newpage

\begin{figure*}[!t]
    \includegraphics[width=\linewidth]{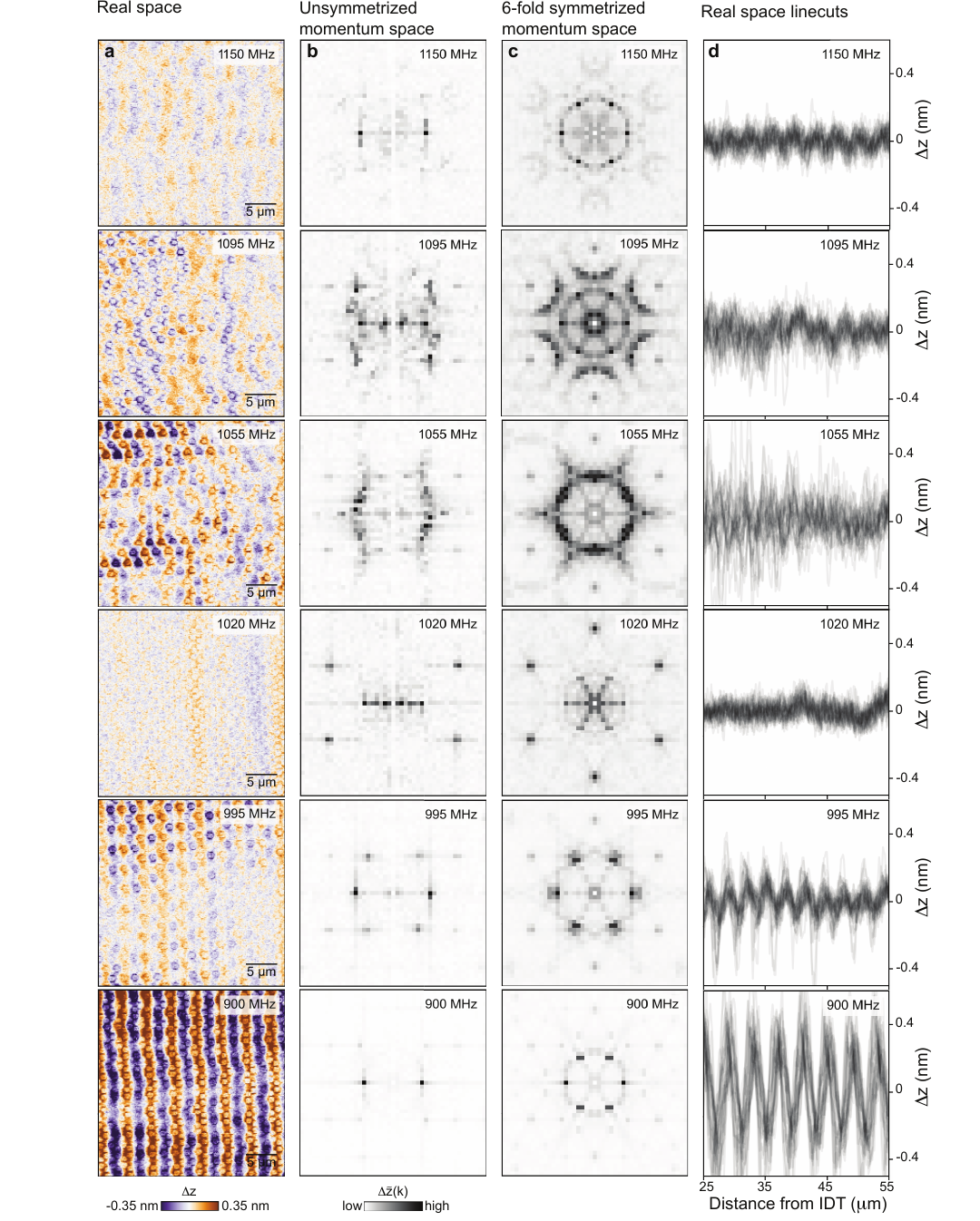}
\caption{\textbf{EFM characterization of the 7.5$\%$ hBN metamaterial.} Same data series as in Fig.~\ref{fig:hbn}, but showing additional frequencies and analysis steps.
\textbf{a,} EFM amplitude shift $\Delta z$ measured at room temperature and ambient pressure for the hBN ($\delta r=7.5\%$) metamaterial. 
A quadratic background was subtracted from each of the 192 scan lines. 
\textbf{b,} Raw Fourier transform of the data in (\textbf{a}). 
\textbf{c,} Six-fold symmetrized data derived from (\textbf{b}). 
\textbf{d,} Overlays of 32 distinct linecuts along the $\hat{x}$ direction from the real-space data in (\textbf{a}), equally spaced in $\hat{y}$.}
\label{fig:hbn_ext}
\end{figure*}

\newpage

\begin{figure*}[!t]
    \includegraphics[width=\linewidth]{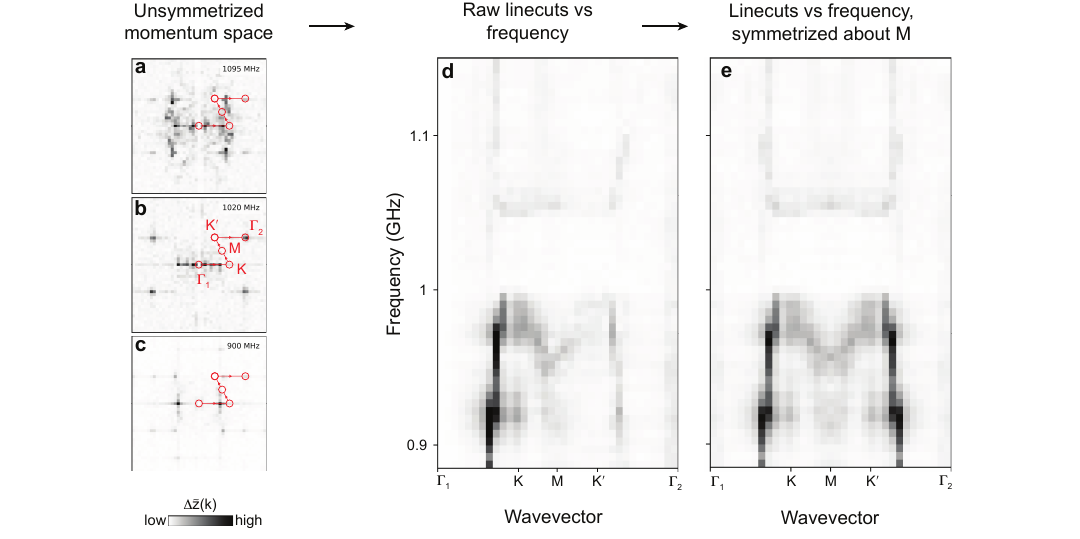}
\caption{\textbf{hBN acoustic bands from unsymmetrized Fourier-space data.}
\textbf{a--c,} Raw Fourier transforms for the hBN ($\delta r=7.5\%$) metamaterial measured above the Dirac frequency $f_{\mathrm{D}}$ (\textbf{a}), at $f_{\mathrm{D}}$ (\textbf{b}), and below $f_{\mathrm{D}}$ (\textbf{c}). 
\textbf{d,} Band dispersion obtained from the $\Gamma_{1}$--K--M--K$^{\prime}$--$\Gamma_{2}$ linecut in momentum space (shown as a dashed red line in (\textbf{b}))  using the data in (\textbf{a}--\textbf{c}).  
\textbf{e,} Band dispersion obtained after symmetrizing the data from (\textbf{d}) about the M point.}
\label{fig:hbn_symm1}
\end{figure*}

\newpage

\begin{figure*}[!t]
    \includegraphics[width=\linewidth]{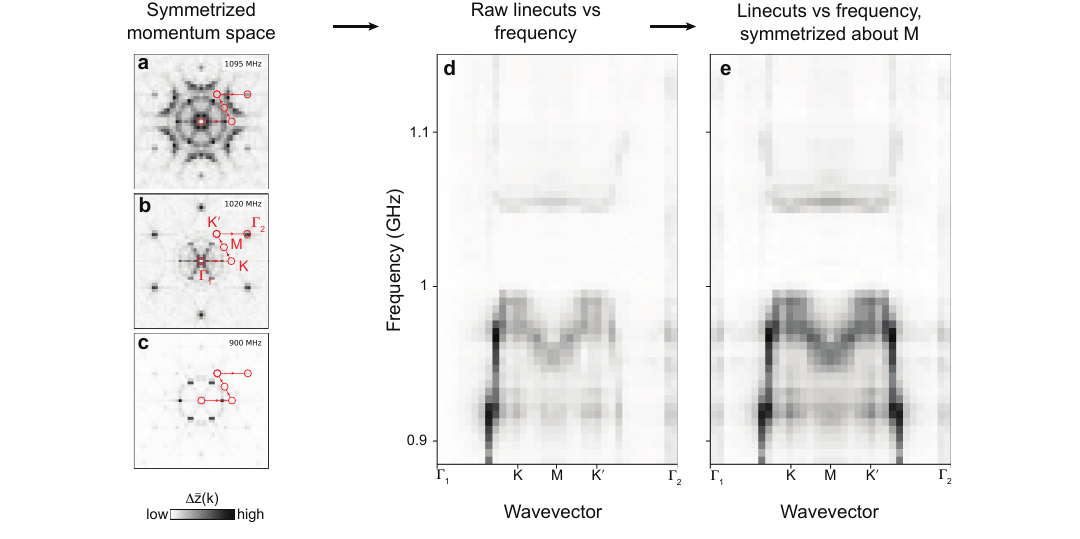}
\caption{\textbf{hBN acoustic bands from symmetrized Fourier-space data.}
\textbf{a--c,} Six-fold symmetrized Fourier transforms for the hBN ($\delta r=7.5\%$) metamaterial measured above the Dirac frequency $f_{\mathrm{D}}$ (\textbf{a}), at $f_{\mathrm{D}}$ (\textbf{b}), and below $f_{\mathrm{D}}$ (\textbf{c}). 
\textbf{d,} Band dispersion obtained from the $\Gamma_1$--K--M--K$^{\prime}$--$\Gamma_{2}$ linecut in momentum space (shown as a dashed red line in (\textbf{b}))  using the data in (\textbf{a}--\textbf{c}).  
\textbf{e,} Band dispersion obtained after symmetrizing the data from (\textbf{d}) about the M point.}
\label{fig:hbn_symm2}
\end{figure*}

\end{document}